# The hypotensive effect of activated apelin receptor is correlated with β-arrestin recruitment

Élie Besserer-Offroy[a,c,ORCID ID], Patrick Bérubé[a,c], Jérôme Côté[a,c], Alexandre Murza[a,c], Jean-Michel Longpré[a,c], Robert Dumaine[a], Olivier Lesur[b,c], Mannix Auger-Messier[b,ORCID ID], Richard Leduc[a,c,ORCID ID], Éric Marsault[a,c,*,ORCID ID], Philippe Sarret[a,c,*]

**Affiliations**

[a] Department of Pharmacology-Physiology, Faculty of Medicine and Health Sciences, Université de Sherbrooke, Sherbrooke, Québec, CANADA J1H 5N4

[b] Department of Medicine, Faculty of Medicine and Health Sciences, Université de Sherbrooke, Sherbrooke, Québec, CANADA J1H 5N4

[c] Institut de pharmacologie de Sherbrooke, Université de Sherbrooke, Sherbrooke, Québec, CANADA J1H 5N4

**e-mail addresses**

Elie.Besserer-Offroy@USherbrooke.ca (ÉBO); Patrick.Berube@USherbrooke.ca (PB); Jerome.Cote@USherbrooke.ca (JC); Alexandre.Murza@USherbrooke.ca (AM); Jean-Michel.Longpre@USherbrooke.ca (JML); Robert.Dumaine@USherbrooke.ca (RD); Olivier.Lesur@USherbrooke.ca (OL); Mannix.Auger-Messier@USherbrooke.ca (MAM); Richard.Leduc@USherbrooke.ca (RL); Eric.Marsault@USherbrooke.ca (EM); Philippe.Sarret@USherbrooke.ca (PS)

**Corresponding Authors**

*To whom correspondence should be addressed:

Philippe Sarret, Ph.D.; Philippe.Sarret@USherbrooke.ca; Tel. +1 (819) 821-8000 ext. 72554

Éric Marsault, Ph.D.; Eric.Marsault@USherbrooke.ca; Tel. +1 (819) 821-8000 ext. 72433

Department of Pharmacology-Physiology, Faculty of Medicine and Health Sciences, Institut de pharmacologie de Sherbrooke, Université de Sherbrooke, Sherbrooke, Québec, CANADA J1H 5N4

---

**Abbreviations used:** 7TMR, seven transmembrane receptor; Akt, protein kinase B; APJ, apelin receptor; βarr, β-arrestin; BRET, bioluminescence resonance energy transfer; cAMP, cyclic adenosine monophosphate; ERK1/2, extra-cellular regulated kinase 1/2; GPCR, G protein-coupled receptor; i.m., intramuscular; i.v., intravenous; MABP, mean arterial blood pressure; MAPK, mitogen-activated protein kinase; NO, nitric oxide; PI3K, phosphoinisitide 3 kinase; PTX, pertussis toxin; RE, relative efficacy; TR-FRET, time-resolved fluorescence resonance energy transfer.

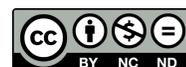

**Abstract**

The apelinergic system is an important player in the regulation of both vascular tone and cardiovascular function, making this physiological system an attractive target for drug development for hypertension, heart failure and ischemic heart disease. Indeed, apelin exerts a positive inotropic effect in humans whilst reducing peripheral vascular resistance. In this study, we investigated the signaling pathways through which apelin exerts its hypotensive action. We synthesized a series of apelin-13 analogs whereby the C-terminal Phe[13] residue was replaced by natural or unnatural amino acids. In HEK293 cells expressing APJ, we evaluated the relative efficacy of these compounds to activate $G\alpha_{i1}$ and $G\alpha_{oA}$ G-proteins, recruit β-arrestins 1 and 2 (βarrs), and inhibit cAMP production. Calculating the transduction ratio for each pathway allowed us to identify several analogs with distinct signaling profiles. Furthermore, we found that these analogs delivered i.v. to Sprague-Dawley rats exerted a wide range of hypotensive responses. Indeed, two compounds lost their ability to lower blood pressure, while other analogs significantly reduced blood pressure as apelin-13. Interestingly, analogs that did not lower blood pressure were less effective at recruiting βarrs. Finally, using Spearman correlations, we established that the hypotensive response was significantly correlated with βarr recruitment but not with G protein-dependent signaling. In conclusion, our results demonstrated that the βarr recruitment potency is involved in the hypotensive efficacy of activated APJ.

**Chemical compounds studied in this article:** [Pyr[1]]-apelin-13 (PubChem CID: 25085173)

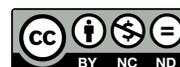

**Graphical Abstract**

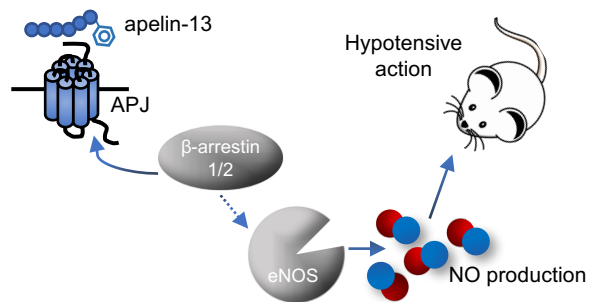

## 1. Introduction

The apelin receptor (angiotensin receptor-like 1, APJ) is a seven transmembrane receptor (7TMR) that belongs to the class A peptidergic G protein-coupled receptors (GPCR) superfamily [1]. The endogenous ligands of APJ are the different isoforms of apelin (namely apelin-13, -17, and -36) [2] as well as the recently discovered ELABELA/Toddler [3]. The apelinergic axis is known to be widely distributed throughout the body, both in the central nervous system and at the periphery [4,5]. Peripherally, components of the apelin/APJ system are mainly expressed in the lungs, kidneys, pancreas as well as in the cardiovascular system [6]. Recently, the apelinergic system has been highlighted as a potential and promising target for drug development [7]. Indeed, modulating APJ signaling can be considered for the treatment of diabetes, obesity, cancer, and HIV infection [4]. However, the most important role of the apelin-APJ axis remains associated to its cardiovascular actions. Indeed, apelin peptides act as potent regulators of vascular function and exert one of the most powerful positive inotropic actions [8]. Accordingly, the apelin-APJ axis is widely distributed within the cardiovascular system, being expressed on endothelial cells, vascular smooth muscle cells and cardiomyocytes [9]. In preclinical models, APJ activation induces a significant drop in mean arterial blood pressure (MABP), reduces ventricular preload and afterload, increases myocardial contractility, and decreases angiotensin II-induced myocardial hypertrophy and fibrosis (for review see [10,11]). In rodents, the hypotensive actions of apelin-13 and apelin-17 are blocked by a pretreatment with L-NAME, a nitric oxide (NO) synthase inhibitor [12]. These beneficial/protective effects of apelin are paralleled in healthy volunteers and in chronic heart failure patients, in whom an intravenous (i.v.) injection of apelin-13 induces a vasodilation of coronary and peripheral blood vessels leading to a lowering of MABP, while increasing the cardiac output [13,14]. In support of the role of the apelin/APJ system in cardiovascular function and pathology, plasma levels of apelin are markedly decreased in patients with chronic heart failure and failing human hearts also exhibit altered apelin and APJ gene expression patterns [15,16].

At the cellular level, activation of APJ triggers several intracellular signaling pathways such as the activation of the $G\alpha_{i/o}$ pathway leading to inhibition of adenylate cyclase and a lowering of cAMP production [5]. Apelin-13 binding to APJ also induces the phosphorylation of the mitogen-activated protein kinase (MAPK) ERK1/2 by a $G\alpha_{i/o}$-dependent mechanism since this effect is blocked by a pretreatment with pertussis toxin (PTX) [17]. APJ activation further induces the recruitment of both β-arrestins 1 and 2 (βarrs), leading to receptor internalization [18,19]. Interestingly, intracellular trafficking routes seem to be agonist-dependent with, on the one hand, apelin-13 and -17, which induce a transient interaction between APJ and βarr, and on the other hand, apelin-36, which displays a long lasting APJ/βarr interaction [20].

Structure-activity relationship (SAR) and alanine-scan have highlighted that the N-terminal amino acids of apelin-13 (i.e. $Arg^2$-$Pro^3$-$Arg^4$-$Leu^5$) are involved in binding to APJ and activation of the $G\alpha_{i/o}$ signaling pathway [5]. However, the C-terminal of apelin plays a crucial role in modulating APJ internalization and βarr recruitment as well as being responsible for the APJ-dependent hypotensive response [18,21]. Indeed, the truncation of the C-terminal Phe residue or its replacement by Ala in apelin-13 or apelin-17 results in an impairment of the MABP lowering effect of these compounds [22]. Accordingly, we recently published that the substitution of $Phe^{13}$ by unnatural amino acids induces changes in cAMP production and also affects the hypotensive action of these C-terminal modified apelin-13 analogs [19]. Herein, we examined the APJ signaling profile of a series of C-terminal modified analogs by studying their ability to activate $G\alpha_{i1}$, $G\alpha_{oA}$, inhibit cAMP production, and recruit βarrs. We further monitored the potency of these compounds to elicit changes in MABP in rodents. To pair APJ receptor activity with its physiological effects, we finally applied the Black and Leff operational model to correlate the drop in MABP with a specific signaling signature.

## 2. Materials and methods

### 2.1. Materials

All compounds including apelin-13 were synthesized by us as previously described [19]. Coelenterazine 400A (DeepBlueC) was purchased from Gold Biotechnology Inc. (St. Louis, MO, USA). DMEM, HEPES (4-(2-hydroxyethyl)-1-piperazineethanesulfonic acid), Penicillin-streptomycin-glutamine and fetal bovine serum (FBS) were obtained from Wisent (St. Bruno, QC, Canada), Opti-MEM was acquired from Invitrogen (Burlington, ON, Canada). Lance Ultra cAMP assay kit was purchased from Perkin Elmer (Montréal, QC, Canada).

### 2.2. Plasmids and constructs

Using an InFusion advantage PCR cloning kit (Clontech Laboratories, Mountain View, CA, USA), the hAPJ construct was inserted into the pIREShygro3-GFP10 vector as previously described [18]. The plasmids encoding RlucII-β-arrestin 1 or 2 [23], GαoA-RlucII [24], Gαi1-RlucII, GFP10-Gγ1, and GFP10-Gγ2 [25] were kindly provided by Dr. Michel Bouvier. The cDNA encoding the human $G_{\beta 1}$ subunit was obtained from the Missouri S&T cDNA Resource Center (Rolla, MO, USA). All constructs were verified by DNA sequencing.

### 2.3. Cell culture and transfections

HEK293 QBI cells (CRL-1573, from ATCC) were cultured in DMEM containing 20 mM HEPES, 10% FBS, penicillin (100 U/mL), streptomycin (100 μg/mL), and glutamine (2 mM) under 5% $CO_2$ at 37°C in a humidified atmosphere. Cells were used between passage numbers 4 to 25. For the transient expression of recombinant proteins, T75 flasks were seeded with $3 \times 10^6$ cells, and 24 h later, the cells were transfected using PEI as previously described [26].

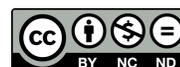

### 2.4. BRET assays

To monitor direct G protein activation, we used the following biosensor couples: $G\alpha_{oA}$-RlucII, GFP10-$G\gamma_1$, and $G\beta1$ [24] or $G\alpha_{i1}$-RlucII, GFP10-$G\gamma_2$, and $G\beta_1$ [25]. G protein biosensors and the hAPJ receptor were transfected into HEK293 QBI cells. At 24 h post-transfection, the cells were detached with trypsin-EDTA and plated (50,000 cells/well) in white opaque 96-well plates (BD Falcon, Corning, NY, USA). At 48 h post-transfection, the cells were washed once with PBS, and 90 µL of HBSS containing 20 mM HEPES was then added. Ligands were added at increasing concentrations for 10 min followed by coelenterazine 400A (5 µM). $BRET^2$ measurements were collected in the 400 to 450 nm window (Rluc) and in the 500 to 550 nm window (GFP10) using the $BRET^2$ filter set on a GENios Pro plate reader (Tecan, Durham, NC, USA). The $BRET^2$ ratio was determined as the light emitted by the acceptor GFP10 over the light emitted by the donor Rluc. The monitoring of βarr recruitment was done by the transient transfection of HEK293 QBI cells with plasmids containing cDNAs encoding hAPJ-GFP10 and Rluc-β-arrestin 1 or 2. The same protocol as the one used for G protein activation was then used.

### 2.5. *Lance Ultra cAMP assay*

The Lance Ultra cAMP assay was performed according to the manufacturer's recommendations as previously described [19]. Briefly, 1,000 cells per well (384-well shallow well plate) were treated with increasing concentrations of compounds for 30 min in the presence of 1 µM forskolin. cAMP-tracer and anti-cAMP-Cryptate were added for at least 1 h. The plates were read on a GENios Pro plate reader with HTRF filters (excitation at 320 nm and emission at 620 and 665 nm). The TR-FRET ratio was determined as the fluorescence of the acceptor (665 nm) over the fluorescence of the donor (620 nm).

### 2.6. *Animal procedures*

All animal procedures were approved by the Ethical and Animal Care Committee of the Université de Sherbrooke and were in accordance with policies and directives of the Canadian Council on Animal Care. Furthermore, all animal studies comply with both the ARRIVE Guidelines (http://www.nc3rs.org.uk/page.asp?id=1357) and the National Institutes of Health guide for the care and use of laboratory animals. Adult male (250-300g) Sprague-Dawley rats (Charles River Laboratories, St-Constant, Québec, Canada) were maintained on a 12 h light/dark cycle with access to food and water *ad libitum*. Rats were acclimatized for 4 days to the animal facility prior to studies.

### 2.6.1. *Blood pressure measurement*

Male Sprague-Dawley rats were anesthetized with a mixture of ketamine/xylazine (87:13 mg/kg i.m.) and placed on a heating pad. Then, systolic, diastolic, and mean arterial pressure were measured every 30 seconds by the tail-cuff method, using the CODA Blood Pressure System (Kent Scientific Co., Connecticut, USA). When the blood pressure was stabilized (five consecutive measure with less than 5% variability), saline, apelin-13, or analogs was injected via the tail vein. Blood pressure measurements were acquired for seven minutes following compound delivery. Data used for Fig.3A were published, in part, in reference [19]. For the present study, the number of rats per dose of apelin-13 was increased and the dose of 10 mg/kg was added in order to fit a dose response curve.

### 2.7. *Data analysis for receptor activation parameters: $EC_{50}$ and $log(\tau/K_A)$*

Each data set was normalized to the maximal response triggered by apelin-13, all values are expressed as the mean ± S.E.M. of at least three different experiments, which were each performed in triplicate. $EC_{50}$ values were obtained from normalized data and determined as the concentration of ligand showing 50% of activation. The data were calculated using concentration-response three parameters non-linear regression of GraphPad Prism 6 (La Jolla, CA, USA).

The Black and Leff operational model describing receptor activation [27] was used to determine the transduction coefficient $\left(\frac{\tau}{K_A}\right)$ for each agonist using eq. (3), which is derived from the standard operational model equation, given by eq. (1):

$$E = \frac{E_m \times [A]^n \times \tau^n}{[A]^n \times \tau^n + ([A] + K_A)^n} \tag{1}$$

Dividing above and below by $K_A$ gave the $\left(\frac{\tau}{K_A}\right)$ ratio, as depicted in eq. (2):

$$E = \frac{E_m \times [A]^n \times \left(\frac{\tau}{K_A}\right)^n}{[A]^n \times \left(\frac{\tau}{K_A}\right)^n + \left(\frac{[A]}{K_A} + 1\right)^n} \tag{2}$$

Dividing the member of eq. (2) by $[A]^n \times \left(\frac{\tau}{K_A}\right)^n$ simplified the equation and gave eq. (3):

$$E = \frac{E_m}{1 + \left(\frac{\frac{[A]}{K_A} + 1}{[A] \times \left(\frac{\tau}{K_A}\right)}\right)^n} \tag{3}$$

In eq. (3), $E$ is the effect triggered by the ligand, $[A]$ is the concentration of ligand, $E_m$ represents the maximum response of the system (this is not the same as the $E_{max}$, which represents the maximal response of the ligand), $n$ is the transducer slope that links occupancy to response (this is not the same as the Hill slope derived from logistic curve fitting, but these two parameters are closely related [28]), $K_A$ depicts the functional equilibrium dissociation constant, and $\tau$ denotes the operational agonist's efficacy. $E_m$ and $n$ are specific of the cell, they are thus shared by all agonists; whereas, $\tau$ and $K_A$ are specific of the ligand.

Fitting experimental data to eq. (3) could result in multiple combinations of $\tau$ and $K_A$, therefore, the transduction coefficient, defined by eq. (4), was used as a parameter to define the agonism for a given pathway at a given receptor:

$$Transduction\ coefficient = log\left(\frac{\tau}{K_A}\right) \tag{4}$$

To eliminate the observational and system bias, ligand activity at a given pathway was divided by the activity of the reference compound, apelin-13 (APE), for the same pathway using eq. (5):

$$\Delta\ log\left(\frac{\tau}{K_A}\right) = log\left(\frac{\tau}{K_A}\right)_{LIG} - log\left(\frac{\tau}{K_A}\right)_{APE} \tag{5}$$

The relative efficacy of an agonist toward a given pathway, relative to the reference compound APE, was calculated as the anti-logarithm of the $\Delta\ log\left(\frac{\tau}{K_A}\right)$ by eq. (6):

$$Relative\ efficacy\ (RE) = 10^{\Delta\ Log\left(\frac{\tau}{K_A}\right)} \tag{6}$$

The standard error of the mean (S.E.M.) of the transduction coefficient $log\left(\frac{\tau}{K_A}\right)$ was calculated by using eq. (7):

$$S.E.M._{log\left(\frac{\tau}{K_A}\right)} = \frac{\sigma}{\sqrt{n}} \tag{7}$$

In eq. (7), $\sigma$ represents the standard deviation calculated from each separate experiment and n is the number of separate experiment performed.

Eq. (8) was used to calculate S.E.M. for the $\Delta log\left(\frac{\tau}{K_A}\right)$ to avoid the propagation of the error during the subtraction step used to calculate the normalized transduction coefficient:

$$S.E.M._{\Delta log\left(\frac{\tau}{K_A}\right)} = \sqrt{\left(S.E.M._{log\left(\frac{\tau}{K_A}\right)_{LIG}}\right)^2 - \left(S.E.M._{log\left(\frac{\tau}{K_A}\right)_{APE}}\right)^2} \quad (8)$$

Standard parameters describing ligand-induced receptor activation (namely $EC_{50}$ and $E_{max}$) are closely related with the Black and Leff operational model of receptor activation. $EC_{50}$ is linked to it by eq. (9):

$$EC_{50} = \frac{K_A}{(2+\tau^n)^{\frac{1}{n}}-1} \quad (9)$$

and $E_{max}$ by eq. (10):

$$E_{max} = \frac{E_m \tau^n}{1+\tau^n} \quad (10)$$

### 2.8. *Data and statistical analyses for the in vivo procedures*

The $ED_{50}$ value of the hypotensive effect of apelin-13 was determined as the concentration of apelin-13 showing 50% of the maximal hypotensive response. Each data point on the dose-response curve represents the maximal hypotensive response of each dose of apelin-13. The data were calculated using dose-response three parameters non-linear regression of GraphPad Prism 6.

Statistical analyses were performed using GraphPad Prism 6 and are described in the figure legend when applicable. A value was considered statistically significant when $p < 0.05$.

## 3. Results and discussion

### 3.1. Chemical modification of the Phe[13] residue results in significant changes in APJ receptor binding and signaling

Before assessing the signaling pathways activated by apelin-13 and its C-terminal modified analogs (structures of the analogs are presented in **Scheme 1**), we first characterized their binding affinities on the human APJ receptor expressed in HEK293 cells. $IC_{50}$ values for each apelin-13 analog are summarized in **Table 1**. The $IC_{50}$ of the endogenous reference compound apelin-13 was 1.2 nM while some tested analogs carrying unnatural amino acids (compounds **3**, **5**, **6**, and **8**) exhibited higher affinities ranging from 0.02 to 0.48 nM. Other C-terminally modified analogs were found to have lower affinities when compared to apelin-13 (compounds **2**, **4**, **7**, **9**, and **10**; with $IC_{50}$ ranging from 6.1 to 68.2 nM). These results are in accordance with previous alanine and D-amino acid scanning data, suggesting that modification of the N-terminal region (i.e. $Arg^2$-$Pro^3$-$Arg^4$-$Leu^5$) as well as C-terminal residue substitutions significantly affect apelin-13 binding to APJ [5,18,19].

To dissect the signaling pathways engaged by this series of apelin-13 analogs, we first assessed the ability of these compounds to activate $G\alpha_{i/o}$ in HEK293 cells transiently overexpressing the APJ receptor. To this end, we used inhibition of the forskolin-induced cAMP production as an endpoint (**Table 1** and **Fig. 1A**). We then monitored the activation of both the $G\alpha_{i1}$ and the $G\alpha_{oA}$ pathways at the direct G protein level. Activation of $G\alpha_{i1}$ by APJ was determined using a BRET-based assay by measuring the dissociation between $G\alpha_{i1}$-RlucII and $G\gamma_2$-GFP10 (**Table 1** and **Fig. 1B**). The same BRET-based cellular approach was used to monitor the activation of $G\alpha_{oA}$ by using the $G\alpha_{oA}$-RlucII and $G\gamma_1$-GFP10 constructs (**Table 1** and **Fig. 1C**) or to measure the ability of these apelin-13 analogs to recruit β-arrestins 1 and 2 (βarrs) to activated APJ-GFP10 receptor (**Table 1** and **Fig. 1D, E**). The use of cells endogenously expressing APJ would have been more relevant, however this was not technically feasible since we need to use the APJ-GFP10 recombinant construct to monitor the recruitment of the Rluc-βarrs at APJ by BRET.

In the cellular functional assays, all compounds acted as full agonists in both the inhibition of cAMP production and $G\alpha_{i/o}$ assays, as they triggered a maximal response ($E_{max}$) comparable to apelin-13 (**Supplementary Table S1**). The potency ($EC_{50}$) of these analogs to inhibit the forskolin-induced cAMP production closely paralleled their binding affinities. Indeed, compounds **6** and **8** were extremely effective at inhibiting cAMP production, with a 20- to 30-fold gain compared to apelin-13. In contrast, compound **7** was found to be more than 100-fold less potent at inhibiting cAMP formation, when compared to apelin-13 (**Table 1**). BRET-based assays monitoring the engagement of $G\alpha_{i1}$ and $G\alpha_{oA}$ revealed potencies in the same order of magnitude as for the reference ligand apelin-13 (**Table 1**). Consequently, we did not observe a direct correlation between cAMP inhibition and $G\alpha_{i/o}$ activation. Compounds **6** and **8** displayed an increase in cAMP inhibition accompanied with no or very small changes in $G\alpha_{i1}$ and $G\alpha_{oA}$ activation. Likewise, compound **7** showed a 5-fold decrease in $G\alpha_{i1}$ activation, no change in $G\alpha_{oA}$ activation while inhibiting cAMP production with 100-fold less potency. In explanation of these discrepancies between cAMP inhibition and G protein activation, it may be argued that C-terminal modified analogs recruit distinct $G\alpha_{i/o}$ isoforms than $G\alpha_{i1}$ and $G\alpha_{oA}$, such as $G\alpha_{i2,3}$ or $G\alpha_{oB}$ to inhibit cAMP production [29]. This hypothesis is supported by the fact that the apelin fragments promote the coupling of APJ to either $G\alpha_{i1}$ or $G\alpha_{i2}$ but not to $G\alpha_{i3}$ [30]. Differences between cAMP inhibition and $G\alpha_{i/o}$ activation may also arise from activation specificities among the nine cloned isoforms of adenylyl cyclase (AC) [31]. Indeed, AC must be pre-stimulated, generally using forskolin to induce cAMP production in order to measure inhibition of the enzyme in a second step. Despite wide use in cell-based high-throughput screening assay, this approach is at risk of biased interpretations since forskolin preferentially activates $AC_I$, $AC_V$ or $AC_{VI}$ and may also bind different adenylyl cyclase regions than G-proteins thus inducing different conformations of the enzyme catalytic core [32-34].

To further investigate the signaling signature of these apelin-13 analogs, we studied βarr recruitment to APJ by measuring BRET signals of Rluc-βarr 1 or 2 association to activated APJ-GFP10 receptor in living cells (**Table 1** and **Fig. 1D, E**). Interestingly, the substitution of the C-terminal $Phe^{13}$ residue of apelin-13 led to different classes of compounds displaying distinct βarr recruitment profiles. Indeed, replacement of $Phe^{13}$ by a lipophilic amino acid such as Val or Ala (compounds **2** and **10**, respectively) decreased the recruitment of both βarrs whereby compound **10** appeared to act as a partial agonist in this paradigm (**Table 1** and **Supplementary Table S1**). Substitution of $Phe^{13}$ with a residue having a shortened side chain residue (PheGly, compound **9**) was also deleterious for recruiting βarrs with a near 5-fold loss compared to the endogenous ligand, apelin-13. On the other hand, replacing $Phe^{13}$ with Phe(4Me), an electron donating residue (compound **3**), or with amino acids having increased electron density and steric hindrance (such as Tyr(OBn) or Bpa, compounds **5** and **6**, respectively), led to an improvement in βarr 1 and βarr 2 recruitment. These results indicate that two physical properties, length and steric hindrance of residue 13's side chain, play a critical role in recruiting βarrs by APJ. We postulate that APJ's orthosteric sub-binding pocket accommodating apelin-13's C-terminal residue needs to be entirely filled to induce βarr recruitment. It has been previously shown that this sub-pocket can accommodate residues with large aromatic side chains but there is a limit of steric hindrance the pocket can tolerate [18]. Indeed, the substitution of Phe[13] with dihydroanthranylAla residue (compound **7**) had a deleterious effect on βarr recruitment. In fact, this compound elicited a 5- and 12-fold loss of βarr 1 and βarr 2 recruitment when compared to apelin-13, respectively (**Table 1**). This negative effect of large side chain residues has been previously observed with a diphenylalanine substitution which led to a 7-fold decrease in βarr 2 recruitment [18]. Furthermore, these findings are supported by the recent published X-ray crystallographic structure of APJ in complex with an apelin-17 derivative showing the shape of the C-terminal cavity delineated by side chains of Y35, W85, Y88, Y93 and Y299 [35]. Finally, the (*L*-α-Me)Phe substitution (compound **8**) was found to improve βarr recruitment compared to apelin-13 (7- and 4-fold increase, respectively) which led to one of the most potent apelin-13 analog reported to date [19].

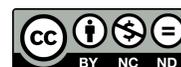

Altogether, these results revealed that modifications of the C-terminal Phe[13] residue exert only small effects on the activation of Gα$_{i/o}$. This is in accordance with previous findings demonstrating that the C-terminal end of apelin-13 and apelin-17 (Lys-Phe-Arg-Arg-apelin-13) does not play a major role in the inhibition of cAMP production [5,12,21]. In sharp contrast, the substitution of the C-terminal Phe[13] residue significantly affects the ability of these modified analogs to recruit β-arrestins. Accordingly, deletion or replacement of the C-terminal residue in apelin-13 or apelin-17 by Ala also influences APJ receptor internalization [12,21,22]. In addition to these *in vitro* observations, the single C-terminal amino acid modification further abolishes the ability of these analogs to modulate blood pressure *in vivo*, thus demonstrating the importance of the role played by this residue in the APJ-triggered MABP lowering [12,36].

### 3.2. Determination of the "within-pathway" relative efficacy of the analogs

Characterization and quantification of the agonist's activity should be evaluated in a manner that allows the use of statistical methods to compare the differences in agonist activity of the tested analogs. Traditional methods, such as comparison of pEC$_{50}$ or E$_{max}$ values, are inadequate due to system bias [37]. To properly quantify the agonistic properties of our analogs, we thus applied the Black and Leff operational model to assess the relative efficacy of our compounds compared to apelin-13 [28]. Considering affinity as well as agonist's efficacy, the Black and Leff [27] operational model specifically quantifies the transduction coefficient defined by Log($\tau$/K$_A$) which fully characterizes the selective activation of cellular signaling pathways by a given agonist. A comprehensive procedure of transduction coefficient quantification is detailed in the methods section and in the supporting information (**Supplementary Note S1**).

Using eq. (3), each concentration-response curve was fitted with the Black and Leff operational model describing receptor activation to derive the transduction coefficient, log($\tau$/K$_A$), as a representation of the analog effect on receptor ability to trigger different signaling pathways (**Supplementary Table S2**). Since apelin-13 is the endogenous ligand of APJ, it was used as the reference ligand to which all other ligands were referred to in a "within-pathway" comparison using eq. (5), resulting in the $\Delta$log($\tau$/K$_A$), the normalized transduction coefficient. The relative efficacy (RE) of the analogs for each pathway, relative to apelin-13, was determined by eq. (6). The normalized transduction coefficient ($\Delta$log($\tau$/K$_A$)) and RE are reported in **Table 2**. A graphical representation of the agonist activity of each analog for all the studied pathways is represented in **Fig. 2**. Compounds **2**, **4**, **7**, **9**, and **10** exhibited similar profiles and were generally less potent than the reference apelin-13 to trigger a cellular response. However, compounds **3**, **5**, **6**, and **8** either showed similar or slightly improved profiles in terms of efficacy when compared to apelin-13. We then quantified the action of our compounds on the APJ-mediated hypotensive action.

### 3.3. Hypotensive effect of apelin-13 and analogs

We first determined the hypotensive action of different doses of the reference ligand apelin-13 by measuring the MABP, in response to a bolus injection, using a non-invasive tail-cuff system in anesthetized male rats. Apelin-13 triggered a rapid and transient MABP drop that returned to basal levels within 6 minutes following intravenous injection (**Fig 3A**). This hypotensive action of apelin-13 was found to be dose-dependent with a determined $ED_{50}$ of $0.12 \pm 0.04$ mg/kg (**Fig. 3B**). We further demonstrated that the hypotensive action of apelin-13 is dependent on the activation of the NO-dependent signaling pathway since L-NAME, a nitric oxide synthase inhibitor, totally abolished apelin-13-triggered MABP drop (**Supplementary Fig. S1**). The acute activation of the apelinergic system observed here is in accordance with previous findings showing that single intravenous injection of apelin-13 in Inactin-anesthetized male Wistar rats led to a rapid blood pressure lowering that last no more than 3-4 min [2,38-40]. Based on these results, it was argued in the literature that the APJ receptor might be a promising therapeutic target in the context of hypertension and pulmonary arterial hypertension [41,42]. Accordingly, the group of Davenport recently demonstrated that daily intraperitoneal injection of ELABELA, a second short-lived endogenous agonist peptide that binds APJ, significantly reduces the pathology of monocrotaline-induced rat model of pulmonary arterial hypertension [43]. Furthermore, acute apelin injection was previously reported to transiently reduce the MABP of spontaneously hypertensive rats almost as efficient as seen in Wistar-Kyoto control rats [44]. Altogether, these results suggest that targeting the apelinergic system might indeed be therapeutically relevant in the context of hypertension.

In order to use a dose that would allow us to detect whether our analogs were more or less potent than apelin-13 to lower the MABP, we then monitored the hypotensive response induced by our compounds at a dose of 0.1 mg/kg, which corresponded to the determined $ED_{50}$ of apelin-13. The maximum drop of MABP measured after iv injection of our analogs is shown in **Fig. 3C**. Compounds **2**, **3**, **5** and **8** were found to reduce MABP, similarly to apelin-13. Compounds **4**, **6**, and **9** triggered an intermediate MABP drop between apelin-13 and saline injections. Furthermore, the ΔMABP of those compounds were not significantly different from apelin-13 or saline injections. Finally, compounds **7** and **10** did not elicit reductions in blood pressure having similar profiles to saline (complete time courses of the blood pressure response for all the analogs are presented in **Supplementary Fig. S2**). These results allowed us to differentiate three classes of compounds: compounds that had similar properties to apelin-13, compounds exhibiting an intermediate profile and compounds that did not induce MABP lowering. We then attempted to see whether a comparable clustering of compounds existed within a particular APJ receptor signaling pathway, thereby correlating physiological effects and cell signaling. Previous results suggested that the MABP drop induced by apelin peptides is mediated by a $G\alpha_{i/o}$-independent signaling pathway but potentially triggered by an internalization-related pathway [12,19,22]. We found that compounds with a MABP reducing behavior similar to apelin-13 usually had a βarr recruitment RE equivalent to that of apelin-13, whereas the intermediate compounds possessed a βarr recruitment RE between 1- and 5-fold less than apelin-13. Finally, compounds **7** and **10** were found to have a βarr recruitment RE of 5- to 15-fold less than the reference. These results prompted us to link βarr recruitment to the hypotensive effect of the compounds. We first assessed whether the normalized transduction coefficients found for all analogs were significantly different from the one of apelin-13 for each of the studied signaling pathways in a Kruskal-Wallis followed by a Dunn's post hoc test (**Supplementary Table S3**). To confirm our hypothesis and since at least one analog possessed a normalized transduction coefficient significantly different from apelin-13 in each signaling assay, we performed different correlations between the ΔMABP and the Δlog($\tau$/KA) for each studied pathway (**Fig. 4** and **Supplementary Fig. S3**). Only the two correlations with the Δlog($\tau$/KA) for βarr recruitment were found to be statistically significant, whereas no correlations were found between ΔMABP and the Δlog($\tau$/KA) for the G$\alpha_{i/o}$-related pathways (i.e. cAMP inhibition, G$\alpha_{i1}$, and G$\alpha_{oA}$). Spearman correlation of ΔMABP and Δlog($\tau$/KA)$_{\beta\text{-arrestin 1}}$ was found to be statistically significant with a *p* value of 0.0022 and a r of 0.867 (**Fig. 4A**). The same correlation with Δlog($\tau$/KA)$_{\beta\text{-arrestin 2}}$ was also found significant with a *p* value of 0.0016 and a r of 0.879 (**Fig. 4B**).

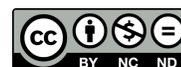

In the present study, we thus demonstrate the key role played by the C-terminal residue of apelin-13 for the modulation of βarr recruitment to APJ and the involvement of this signaling pathway in the hypotensive response mediated through APJ. We also found that modulation of the βarr recruitment by substituting Phe[13] with natural and unnatural amino acids led to the modulation of the blood pressure lowering effect. This is also confirmed by a recent study aimed at developing apelin-13 macrocyclic analogs where three compounds (compounds **7**, **15** and **20**; structures of these compounds can be found in **Supplementary Scheme S1**, upper panel) were shown to induce a hypotensive effect significantly different from the one mediated by apelin-13 at an equimolar dose [45]. When looking at the activation profile of those analogs, we further observed that those three compounds had a decreased ability to induce βarr recruitment compared to apelin-13 while they conserved their capacity to bind to APJ and to activate $G\alpha_{i1}$ and $G\alpha_{oA}$ pathways. Likewise, the cyclic apelin peptide MM07, which preferentially activates G-protein coupling over βarr signaling was reported to be not effective in lowering MABP [46].

Furthermore, this structure-activity relationship (SAR) was also true when assessing the second endogenous ligand of APJ, ELABELA/Toddler. Indeed, the discovery and SAR of the minimally bioactive fragment of ELABELA/Toddler revealed two compounds (compounds **3** and **4**; see compound's structures in **Supplementary Scheme S1**, lower panel) that displayed biased profiling comparable to the one observed for the macrocyclic compounds [47]. Those two compounds exhibited lower efficacies at recruiting βarr 2 and lower reduction of MABP. Altogether, our results, as well as previously published findings indicate that βarr recruitment at the APJ receptor represents the initiating signaling step that regulates the hypotensive action mediated by apelin peptides. Interestingly, ELA(19-32), compound **3**, still triggers an increased left ventricular developed pressure (LVDP) in the same range of apelin-13 while its potency to mediate a hypotensive response is decreased. This suggests that the inotropic effect of APJ is mediated through a signaling pathway distinct from βarr recruitment [8].

It has already been proposed that βarr recruitment plays an important role in blood vessel relaxation. Indeed, apelin-17 was shown to counteract angiotensin II mediated glomerular arteriole constriction while apelin-17 lacking the C-terminus amino acid (K16P) had no effect on arteriole diameter [22]. The authors thus concluded that apelin-17 caused a decrease in MABP through a βarr-dependent ERK1/2 activation process. Following binding to APJ, apelin-13 has, however, been shown to provoke PTX-sensitive ERK1/2 phosphorylation via a PKC-dependent mechanism in a Ras-independent manner [17]. It is therefore unlikely here that the hypotensive action of apelin-13, mediated via the βarr-dependent pathway, relies on the activation of the ERK1/2 signaling cascade. This is not the first time that the binding of different apelin fragments (i.e. apelin-13, -17 or -36) to the APJ receptor behaves differently. Lee *et al.*, have previously reported that apelin-13 induces a rapid recycling of the receptor to the cell surface after internalization via a Rab4-dependent process, whereas apelin-36 internalized receptors remain associated with βarrestins and are targeted to the lysosomal compartment by Rab7 [20]. Likewise, it was also demonstrated that short and long forms of apelin exhibit distinct desensitization profiles. Indeed, the deletion of the C-terminal tail of APJ decreases the desensitization pattern induced by apelin-36 without altering the desensitization profile related to APJ activation by apelin-13 [30].

The hypotensive action of apelin has been shown to be mediated through a NO-dependent mechanism since L-NAME blocked the apelin-mediated MABP lowering. [12,48]. Therefore, the link between βarr recruitment and NO production needs to be addressed. The serine/threonine kinase, Akt, has been shown to be activated by several proteins such as PI3K and βarrs [49-51] and has also been linked to the activation of the endothelial isoform of NO-synthase [52]. We can therefore hypothesize that activation of Akt could stimulate NO production in endothelial cells by a βarr-dependent mechanism. This is of particular interest since apelin, following binding on APJ, was shown to induce the phosphorylation of Akt in endothelial cells [53,54]. Furthermore, mechanical stretch has been recently reported to activate APJ in a biased manner by favoring βarr over $G\alpha_{i/o}$ signaling [55,10]. Thus, this mechanism could play a protective role by inducing a hypotensive response, hence decreasing mechanical stretch.

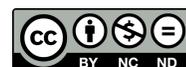

## 4. Conclusion

We have demonstrated that the hypotensive effect mediated by APJ is significantly correlated with
the recruitment of βarrs. Furthermore, we identified that the recruitment of βarr can be modulated
by modifying the C-terminal residue of apelin-13 and that potent agonists at stimulating βarr
translocation were effective in lowering MABP. We bring new insight into APJ-mediated
signaling and the relationship that exists between the physiological action of a hormone and its
downstream signaling pathways. Further studies are, however, required to discriminate which
signaling pathway, acting downstream of βarr activation, is involved in the hypotensive action
mediated by apelin binding to APJ. Finally, the development of biased apelin receptor agonists
will be of great interest to decipher the physiopathological roles played by the apelinergic axis and
to potentially offer new and effective treatment options for cardiovascular disease.

## 5. Supplementary material

Fitting parameters and equations to retrieve ligand-specific parameters of receptor activation using
GraphPad Prism, structure of cited macrocyclic analogs of apelin-13 and ELABELA analogs,
effect of L-NAME on apelin-13-mediated hypotensive response, time-course of the hypotensive
response triggered by the apelin analogs, correlations between ΔMABP and the $\Delta\log(\tau/K_A)$ for
the $G\alpha_{i/o}$-related pathways, Emax table, $\log(\tau/K_A)$ table, and statistical analysis of $\Delta\log(\tau/K_A)$
table are available as supplementary information PDF file.

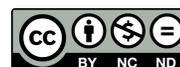

## Acknowledgments

ÉBO was supported by a research fellowship from the Institut de pharmacologie de Sherbrooke (IPS) and Centre d'excellence en neurosciences de l'Université de Sherbrooke (CNS). PS holds a Canada Research Chair in Neurophysiopharmacology of Chronic Pain. Drs M. Bouvier, T. Hebert, S.A. Laporte, G. Pineyro, J.-C. Tardif and E. Thorin (CQDM Team) are also acknowledged for providing us with the G-protein BRET-based biosensors.

## *Funding*

This work was supported by the Canadian Institute of Health Research (CIHR) [grant number FDN-148413] to PS, the National Science and Engineering Research Council of Canada (NSERC) [grant number CRD-399680] to ÉM, and the FRQ-S funded Réseau québécois de recherche sur le médicament (RQRM). The authors declare no competing financial interests.

## *Authors contribution*

Conception and design of study: ÉBO, ÉM, and PS

Acquisition of data: ÉBO, JML, PB, and JC

Analysis and interpretation of data: ÉBO, JML, ÉM, and PS

Contributed new reagent and analysis tools: ÉBO, AM, and ÉM

Wrote the manuscript: ÉBO, RD, OL, MAM, RL, JML, ÉM, and PS

All the authors approved the version of the manuscript to be published.

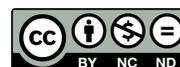

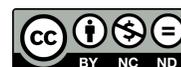

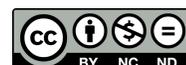

## Schemes, Figures, and Tables

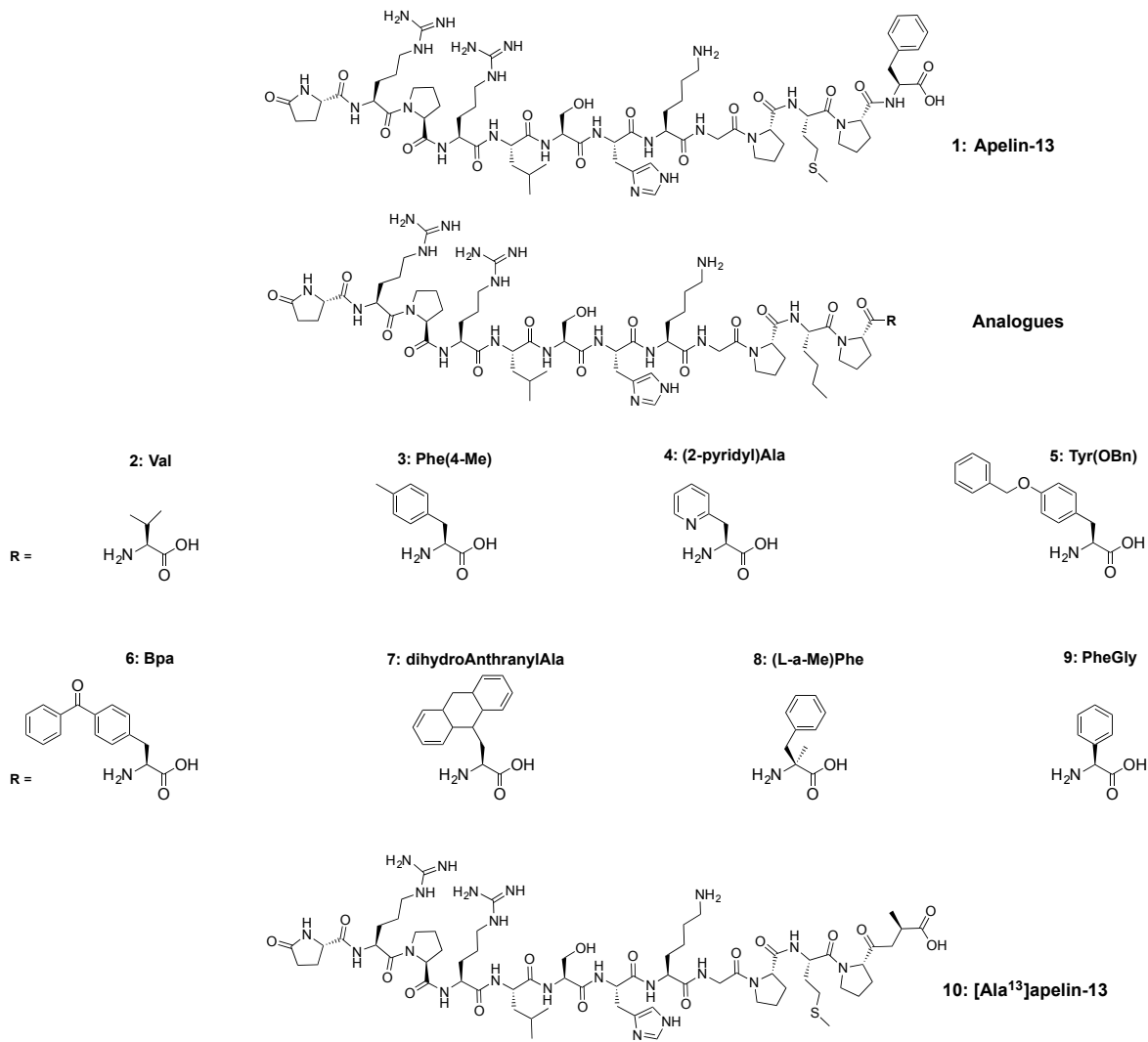

Scheme 1. Developed structures of apelin-13 and analogs used in this study.

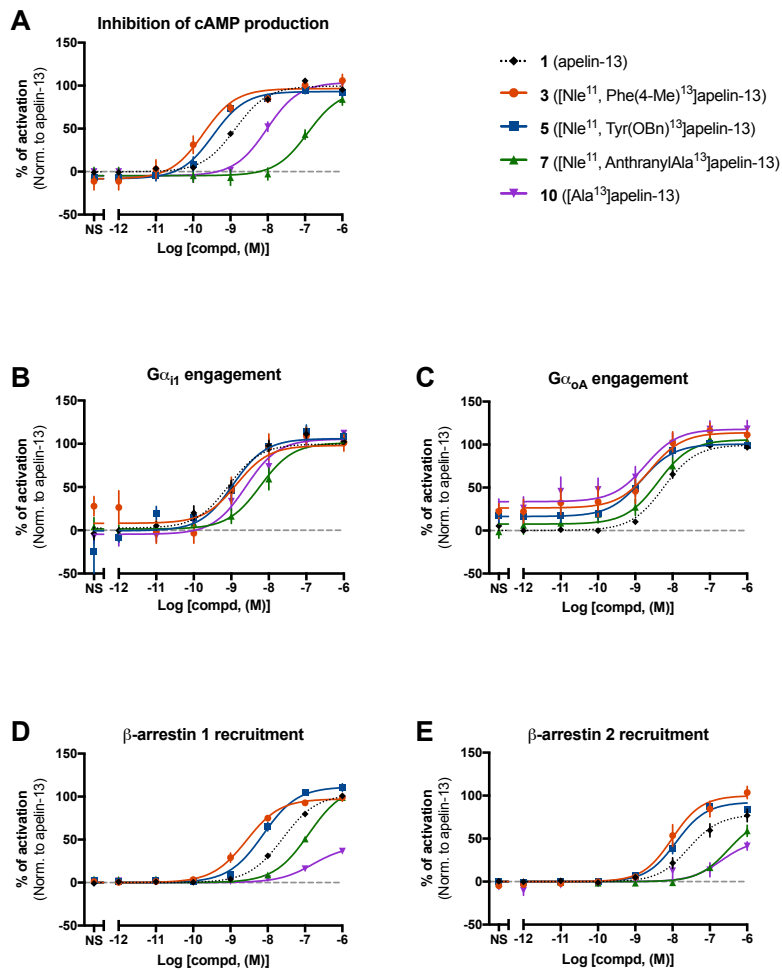

**Fig. 1**. **Activation of signaling pathways triggered by apelin-13 and four representative C-terminally modified analogs**. (**A**) Ligand-mediated inhibition of the forskolin-stimulated accumulation of cAMP measured using PerkinElmer's Lance Ultra cAMP assay. (**B-C**) Ligand-triggered engagement of the G protein pathways $G\alpha_{i1}$ (**B**) and $G\alpha_{oA}$ (**C**) monitored using the BRET-based G protein dissociation assay. (**D-E**) Ligand-induced recruitment of βarr 1 (**D**) and βarr 2 (**E**) using the BRET-based βarr recruitment assay. Each data set represents the mean of three independent experiments, each performed in triplicate, and expressed as the mean ± S.E.M.

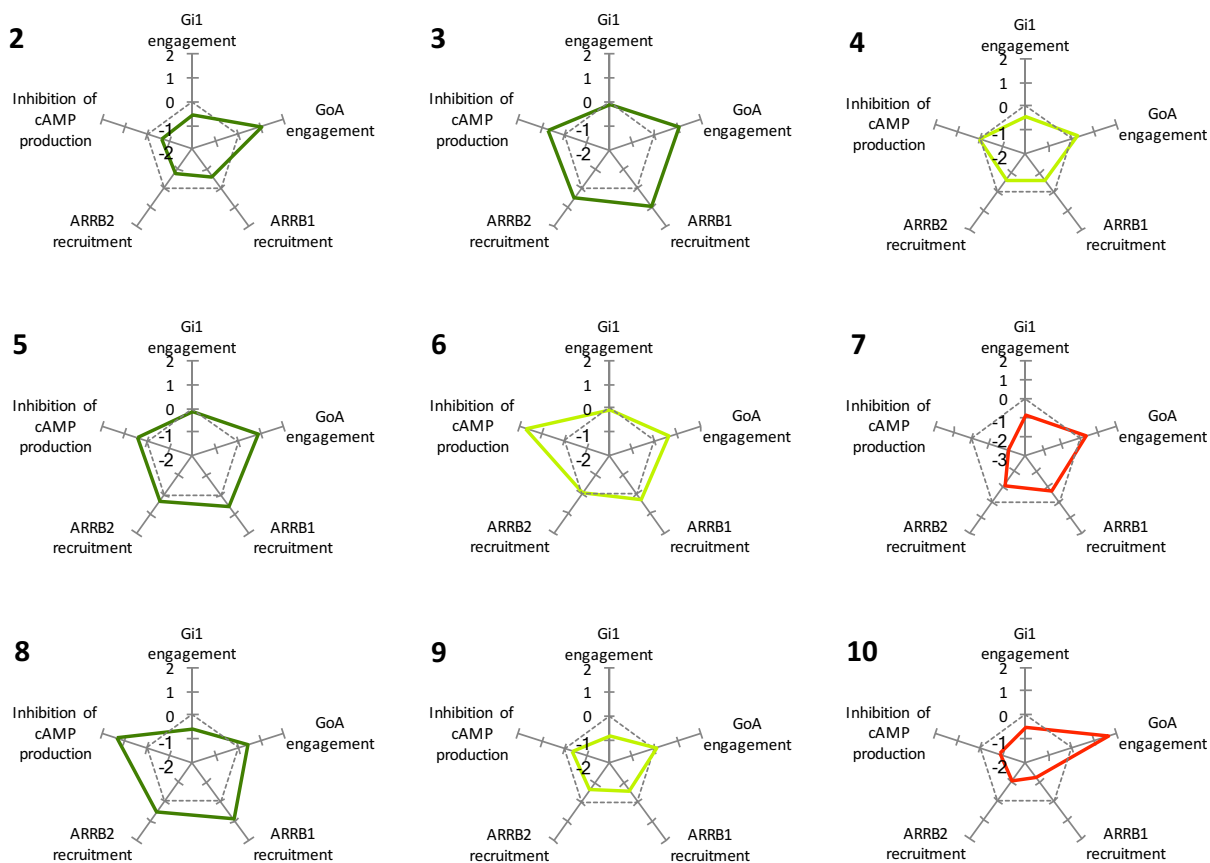

**Fig. 2. Relative efficacy plots for each analog**. The normalized transduction coefficient ($\Delta\log(\tau/K_A)$), as calculated using the Black and Leff operational model, was graphed for each studied signaling pathway. Values for each analog are represented by the solid colored line and apelin-13's values are represented on each graph by the gray dashed line. Each point on graphs represents the mean $\Delta\log(\tau/K_A)$ of the analog for the specified signaling pathway and was calculated using concentration-response curves, which were the result of three independent experiments.

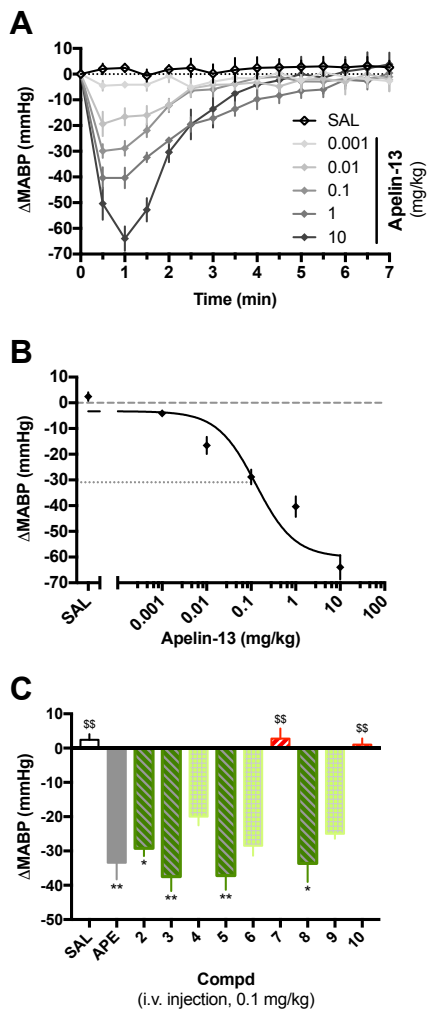

**Fig. 3. Effects of apelin-13 and C-terminally modified analogs on blood pressure**. (**A**) Effect of increasing doses of apelin-13 (i.v. injection) on the mean arterial blood pressure (MABP) of anesthetized rats, measured using non-invasive blood pressure recording. (**B**) Dose-response curve of apelin-13's hypotensive effects. Each point represents the maximum blood pressure drop as recorded on panel A. (**C**) Effect of the i.v. injection of saline (SAL, 0.1 mL/kg), apelin-13 (APE) or the analogs (0.1 mg/kg) on the MABP of anesthetized rats. Each value represents the mean ± S.E.M. obtained with 6-8 rats. Statistical analysis was performed on the ΔMABP values using the Kruskal-Wallis multiple comparisons followed by a Dunn's post-hoc test. * $p < 0.05$, ** $p < 0.01$ compared to saline; \$\$ $p < 0.01$ compared to apelin-13.

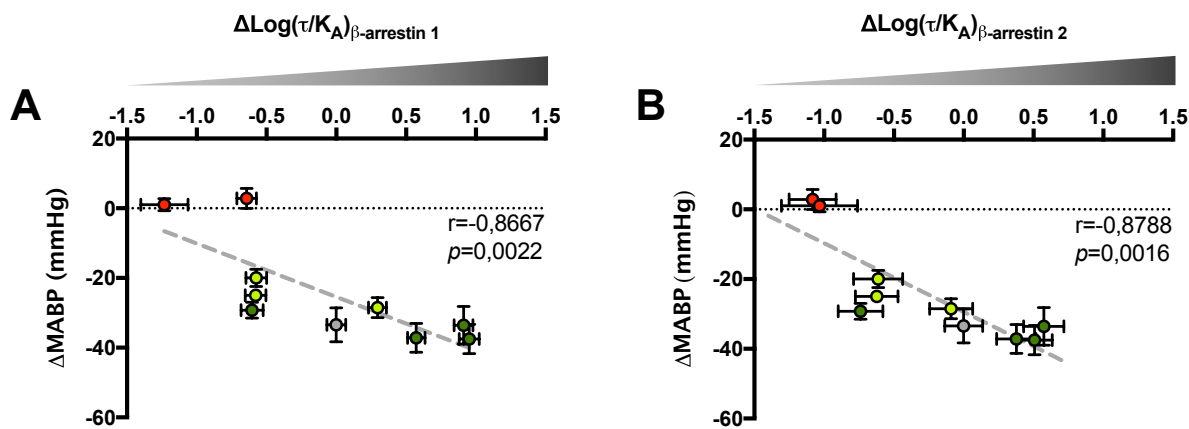

**Fig. 4. Correlations between the hypotensive effect and the relative efficacy of βarr recruitment.** (**A**) Correlation between the ΔMABP and the normalized transduction coefficient for βarr 1 recruitment, denoted by the $\Delta\log(\tau/K_A)_{\beta\text{-arrestin 1}}$. (**B**) Correlation between the ΔMABP and the normalized transduction coefficient for βarr 2 recruitment, denoted by the $\Delta\log(\tau/K_A)_{\beta\text{-arrestin 2}}$. Each point represents the mean ± S.E.M. of the ΔMABP in the y-axis and the S.E.M. of the $\Delta\log(\tau/K_A)$ in the x-axis. Colors denote the ability to induce a hypotensive response, red dots represent analogs statistically different from apelin-13, light green dots represent analogs not statistically different from saline or apelin-13, and dark green dots represent analogs statistically different from saline. The gray dot represents the reference compound apelin-13 as represented in Fig. 3C. Statistical analysis was performed using the Spearman correlation; r coefficient and *p* values are reported on each graph, a *p* < 0.05 was considered significant.

| Analog | sequence | binding IC$_{50}$ (nM)[a] | functional assays[b] | | | | |
|---|---|---|---|---|---|---|---|
| | | | cAMP$_i$ EC$_{50}$ (nM)[c] | G$\alpha_{i1}$ EC$_{50}$ (nM) | G$\alpha_{oA}$ EC$_{50}$ (nM) | $\beta$arr 1 EC$_{50}$ (nM) | $\beta$arr 2 EC$_{50}$ (nM) |
| **1** | apelin-13 | 1.2 ± 0.1 | 1.8 ± 0.4 | 0.39 ± 0.07 | 6.2 ± 1.4 | 24 ± 1.4 | 52 ± 13 |
| **2** | [Nle$^{11}$, Val$^{13}$]apelin-13 | 18.2 ± 3.2 | 6.9 ± 3.4 | 0.52 ± 0.03 | 3.4 ± 1.1 | 81 ± 3 | 232 ± 99 |
| **3** | [Nle$^{11}$, Phe(4-Me)$^{13}$]apelin-13 | 0.25 ± 0.02 | 0.13 ± 0.01 | 0.56 ± 0.02 | 3.0 ± 1.5 | 3.4 ± 1.5 | 33 ± 26 |
| **4** | [Nle$^{11}$, (2-pyridyl)Ala$^{13}$]apelin-13 | 6.1 ± 1.0 | 4.8 ± 2.5 | 0.32 ± 0.05 | 2.2 ± 0.6 | 18 ± 5.9 | 35 ± 6.8 |
| **5** | [Nle$^{11}$, Tyr(OBn)$^{13}$]apelin-13 | 0.02 ± 0.003 | 0.31 ± 0.16 | 0.49 ± 0.21 | 1.5 ± 0.5 | 7.2 ± 2.6 | 14 ± 5.5 |
| **6** | [Nle$^{11}$, Bpa$^{13}$]apelin-13 | 0.48 ± 0.05 | 0.06 ± 0.01 | 0.64 ± 0.30 | 4.0 ± 2.0 | 13 ± 3.4 | 38 ± 10 |
| **7** | [Nle$^{11}$, AnthranylAla$^{13}$]apelin-13 | 13.7 ± 1.6 | 217 ± 51 | 2.14 ± 0.18 | 5.1 ± 1.3 | 133 ± 10 | 628 ± 223 |
| **8** | [Nle$^{11}$, (*L*-$\alpha$-Me)Phe$^{13}$]apelin-13 | 0.43 ± 0.03 | 0.09 ± 0.03 | 2.32 ± 0.85 | 3.6 ± 1.5 | 3.2 ± 1.8 | 15 ± 7.0 |
| **9** | [Nle$^{11}$, PheGly$^{13}$]apelin-13 | 14.3 ± 1.4 | 2.4 ± 0.4 | 1.62 ± 0.20 | 10.3 ± 3.9 | 102 ± 24 | 247 ± 160 |
| **10** | [Ala$^{13}$]apelin-13 | 68.2 ± 15 | 8.84 ± 1.3 | 2.41 ± 0.97 | 1.9 ± 0.6 | 425 ± 303 | 154 ± 98 |

[a]*Values of binding are from* [19]. *IC$_{50}$ values represent the concentration of ligand inhibiting 50% of the radioligand binding.* [b]*EC$_{50}$ values represent the concentration of ligand showing 50% of pathway's activation.* [c]*cAMP$_i$ refers to the inhibition of the forskolin-induced cAMP production.*

**Table 1. Binding affinity and Gα$_{i/o}$, cAMP inhibition, and βarr recruitment potencies of C-terminally modified apelin-13 analogs**.

Each value represents the mean of three independent experiments, which were each done in triplicate, and is expressed as the mean ± S.E.M.

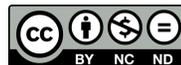

| Analog | cAMP$_i$[a] | | G$\alpha_{i1}$ | | G$\alpha_{oA}$ | | $\beta$arr 1 | | $\beta$arr 2 | |
|---|---|---|---|---|---|---|---|---|---|---|
| | $\Delta\log(\tau/K_A)$ | RE | $\Delta\log(\tau/K_A)$ | RE | $\Delta\log(\tau/K_A)$ | RE | $\Delta\log(\tau/K_A)$ | RE | $\Delta\log(\tau/K_A)$ | RE |
| **1** | 0.00 ± 0.07 | 1.00 | 0.00 ± 0.22 | 1.00 | 0.00 ± 0.26 | 1.00 | 0.00 ± 0.07 | 1.00 | 0.00 ± 0.14 | 1.00 |
| **2** | -0.61 ± 0.11 | 0.25 | -0.60 ± 0.25 | 0.25 | 1.01 ± 0.26 | 10.26 | -0.60 ± 0.08 | 0.25 | -0.74 ± 0.16 | 0.18 |
| **3** | 0.68 ± 0.11 | 4.81 | -0.16 ± 0.24 | 0.70 | 1.09 ± 0.26 | 12.39 | 0.95 ± 0.07 | 8.95 | 0.51 ± 0.13 | 3.22 |
| **4** | 0.05 ± 0.14 | 1.12 | -0.45 ± 0.23 | 0.36 | 0.32 ± 0.25 | 2.09 | -0.57 ± 0.07 | 0.27 | -0.61 ± 0.18 | 0.24 |
| **5** | 0.45 ± 0.11 | 2.85 | -0.14 ± 0.22 | 0.72 | 0.91 ± 0.26 | 8.09 | 0.57 ± 0.06 | 3.74 | 0.38 ± 0.14 | 2.38 |
| **6** | 1.71 ± 0.09 | 51.40 | -0.09 ± 0.22 | 0.81 | 0.62 ± 0.27 | 4.14 | 0.30 ± 0.06 | 1.97 | -0.09 ± 0.15 | 0.81 |
| **7** | -2.09 ± 0.14 | 0.01 | -0.86 ± 0.24 | 0.14 | 0.38 ± 0.26 | 2.40 | -0.64 ± 0.07 | 0.23 | -1.08 ± 0.17 | 0.08 |
| **8** | 1.37 ± 0.12 | 23.5 | -0.62 ± 0.25 | 0.24 | 0.46 ± 0.26 | 2.88 | 0.91 ± 0.07 | 8.17 | 0.57 ± 0.14 | 3.75 |
| **9** | -0.32 ± 0.11 | 0.47 | -0.87 ± 0.23 | 0.13 | 0.04 ± 0.25 | 1.10 | -0.58 ± 0.07 | 0.26 | -0.62 ± 0.15 | 0.24 |
| **10** | -0.86 ± 0.10 | 0.14 | -0.54 ± 0.29 | 0.29 | 1.64 ± 0.26 | 43.75 | -1.23 ± 0.17 | 0.06 | -1.03 ± 0.27 | 0.09 |

[a]*cAMP$_i$ refers to the inhibition of the forskolin-induced cAMP production.*

**Table 2. Normalized transduction coefficient ($\Delta\log(\tau/K_A)$) and relative efficacy (RE) of C-terminally modified apelin-13 analogs**.

Each value represents the mean of three independent experiments, done in triplicate, and expressed as the mean ± S.E.M.

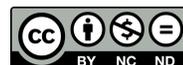

# The hypotensive effect of activated apelin receptor is correlated with β-arrestin recruitment

## *Supporting Information*

Élie Besserer-Offroy[a,c],ORCID ID, Patrick Bérubé[a,c], Jérôme Côté[a,c], Alexandre Murza[a,c], Jean-Michel Longpré[a,c], Robert Dumaine[a], Olivier Lesur[b,c], Mannix Auger-Messier[b],ORCID ID, Richard Leduc[a,c],ORCID ID, Éric Marsault[a,c],*,ORCID ID, Philippe Sarret[a,c]*

**Affiliations**

[a] Department of Pharmacology-Physiology, Faculty of Medicine and Health Sciences, Université de Sherbrooke, Sherbrooke, Québec, CANADA J1H 5N4

[b] Department of Medicine, Faculty of Medicine and Health Sciences, Université de Sherbrooke, Sherbrooke, Québec, CANADA J1H 5N4

[c] Institut de pharmacologie de Sherbrooke, Université de Sherbrooke, Sherbrooke, Québec, CANADA J1H 5N4

*To whom correspondence should be addressed

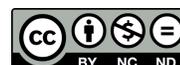

**TABLE OF CONTENT**

**Supplementary Note S1. Fittings parameters and equations to retrieve ligand-specific parameters of receptor activation using GraphPad Prism.**

Selection of the reference agonist: Any ligand could be defined as the reference agonist providing it is a full agonist on each of the assessed pathways.

The following code is used to program a non-linear curve fit to retrieve ligand-specific parameters of receptor activation with GraphPad Prism v6 and higher:

1. Open GraphPad Prism and create a new user-define equation:

    Go to *Analyse* 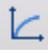 > *Create new equation*

2. In the *equation tab*, select "*Explicit Equation: Y = a function of X and parameters*"

3. Name your new equation: "*Black and Leff operational model*"

4. In the *definition panel* enter the following:

    A_T=10^(X*n)*10^(LogT*n)

    Y=Emax*A_T/[A_T+(10^X+10^LogKA)^n]

5. Go to the "*Rules for Initial Values*" tab and enter the following:

| Parameter | Value | Rule |
|:---:|:---:|:---:|
| n | 2 | Initial value, to be fit |
| LogT | 0.1 | Initial value, to be fit |
| Emax | 100 | Initial value, to be fit |
| LogKA | -8 | Initial value, to be fit |

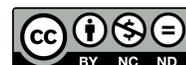

Select: *Start graphing the curve at the smallest X value*

6.  Go to the "*Default Constraints*" tab and enter the following:

| Parameter | Constraint | Value |
|:---:|:---:|:---:|
| n | Shared value for all dataset | - |
| LogT | No constraint | - |
| Emax | Constant equal to | 100 |
| LogKA | No constraint | - |

7.  Go to the "*Default Constraints*" tab and enter the following:

    a.  In "*Report transforms of best-fit parameters (with 95% CI)*":

       KA = 10^LogKA          CI method: Asymetrical

       T = 10^LogT            CI method: Asymetrical

    b.  In "*Report these combinations of best-fit parameters (with 95% CI)*":

       Log(T/KA) = P1 - P2      P1 = LogT       P2 = LogKA

8.  Your new custom equation is now configured. You can retrieve it under:

*Analyse > Nonlinear regression > User-defined equation > Black and Leff operational model*

9.  To analyze your data, create a new XY spreadsheet and enter the normalized values from your assay into column A and subsequent. Then, retrieve your custom equation and apply the analysis.

10. If curve fit does not work (Ambigous), modify the Emax constraint and enter the Emax of the reference ligand.

11. The Log(T/KA) is given directly in "*results of nonlinear fit*".

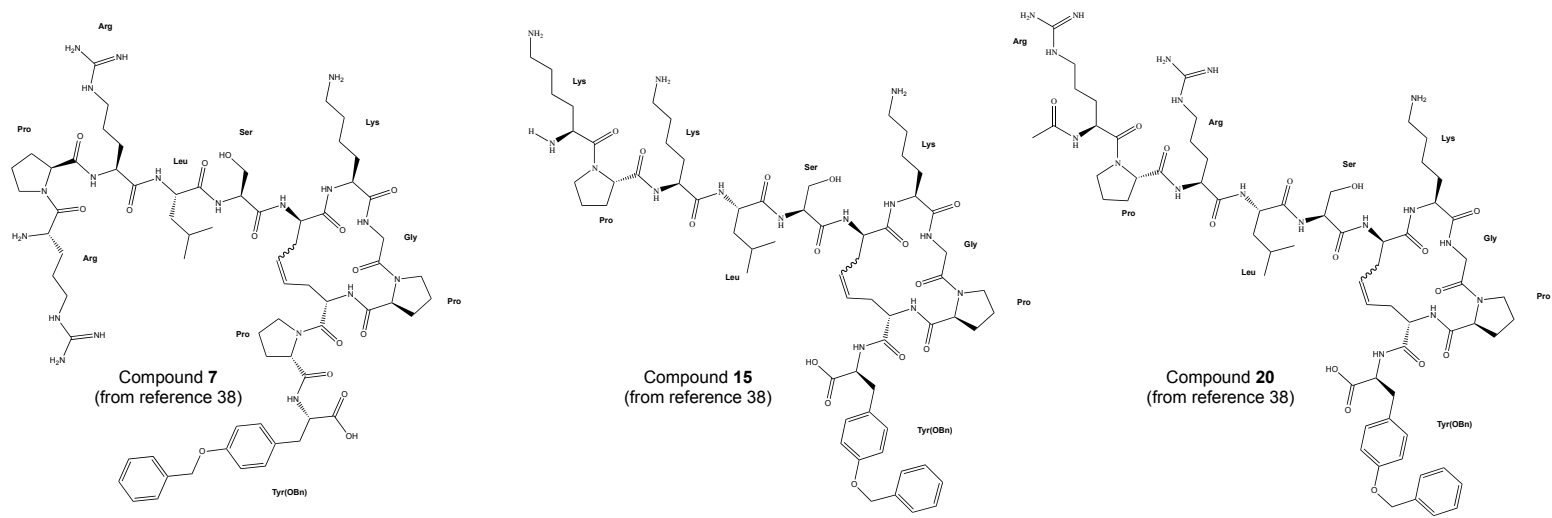

Compound **7**
(from reference 38)

Compound **15**
(from reference 38)

Compound **20**
(from reference 38)

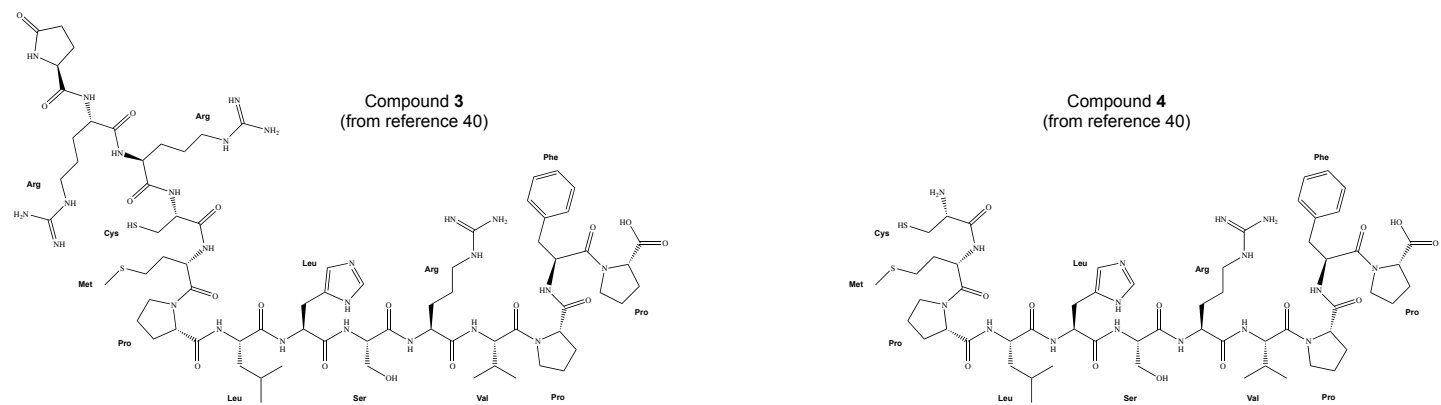

Compound **3**
(from reference 40)

Compound **4**
(from reference 40)

**Supplementary Scheme S2. Developed formulae of compounds 7, 15, and 20 (from the reference 38) and developed formulae of compounds 3 and 4 (from the reference 40).**

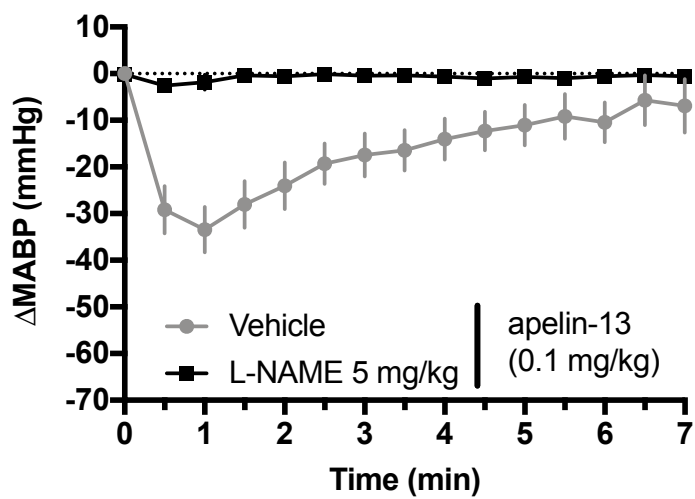

**Supplementary Fig. S1. Time course of the blood pressure response triggered by the bolus injection of apelin-13 in L-NAME- or vehicle-treated rats.** The grey solid line represents the effect of a 0.1 mg/kg dose of apelin-13 injected as a bolus in vehicle-treated animals. The black (filled squares) line represents the effect of a bolus injection of apelin-13 at a dose of 0.1 mg/kg in animals pre-treated with 5 mg/kg of the NO-synthase inhibitor L-NAME. Bolus injection of apelin-13 was done 15 minutes after vehicle or L-NAME treatment. Each point represents the mean ± S.E.M. obtained with 5-8 rats.

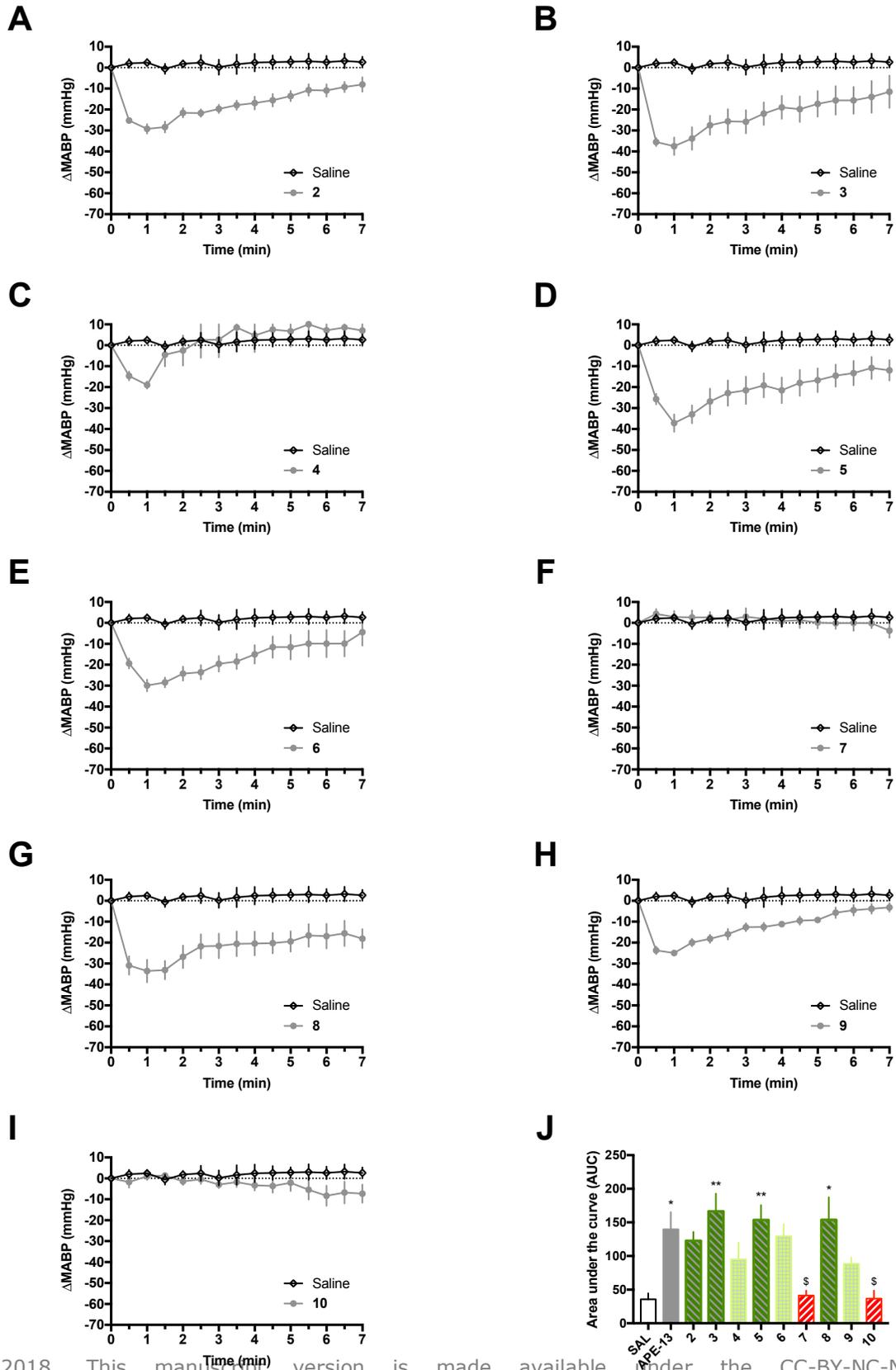

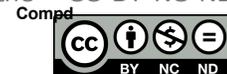

**Supplementary Fig. S2. Time course of the blood pressure response triggered by the bolus injection of apelin-13 analogs.** (**A**) Compound **2**. (**B**) Compound **3**. (**C**) Compound **4**. (**D**) Compound **5**. (**E**) Compound **6**. (**F**) Compound **7**. (**G**) Compound **8**. (**H**) Compound **9**. (**I**) Compound **10**. The black solid line represents the effect of saline injection. (**J**) Area under the curve for saline injection (SAL), apelin-13 (APE-13), and each of the analogs (**2-10**). Each point represents the mean ± S.E.M. obtained with 6-8 rats. Statistical analysis was performed on the AUC values using the Kruskal-Wallis multiple comparisons followed by a Dunn's post-hoc test. * $p < 0.05$, ** $p < 0.01$ compared to saline; \$ $p < 0.05$ compared to apelin-13.

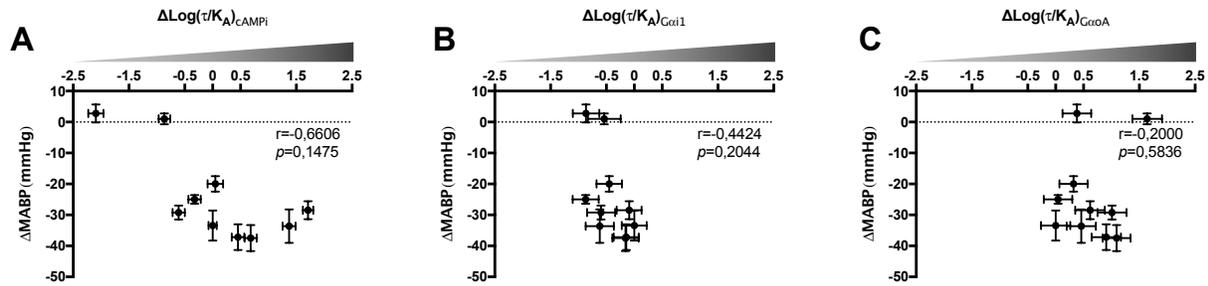

**Supplementary Fig. S3. Correlations between the hypotensive effect and the efficacy to activate Gα$_{i/o}$ and cAMP signaling pathways.** (**A**) Correlation between the ΔMABP and the normalized transduction coefficient for the inhibition of cAMP production, denoted by the Δlog(τ/KA)$_{cAMPi}$. (**B**) Correlation between the ΔMABP and the normalized transduction coefficient for the activation of the Gα$_{i1}$ pathway, denoted by the Δlog(τ/KA)$_{Gαi1}$. (**C**) Correlation between the ΔMABP and the normalized transduction coefficient for the activation of the Gα$_{oA}$ pathway, denoted by the Δlog(τ/KA)$_{GαoA}$. Each point represents the mean ± S.E.M. of the ΔMABP in the y-axis and the S.E.M. of the Δlog(τ/KA) in the x-axis. Statistical analyses were performed using the Spearman correlation; r coefficient and p values are reported on each graph.

**Supplementary Table S1. Maximal efficacy ($E_{max}$) values of each compound on $G\alpha_{i/o}$, cAMP inhibition, and β-arrestin recruitment.** Each value represents the mean of three independent experiments, which were each done in triplicate, and is expressed as the mean ± S.E.M.

| analog | $cAMP_i$[a] $E_{max}$ (%) | $G\alpha_{i1}$ $E_{max}$ (%) | $G\alpha_{oA}$ $E_{max}$ (%) | β-arrestin 1 $E_{max}$ (%) | β-arrestin 2 $E_{max}$ (%) |
|---|---|---|---|---|---|
| **1** | 100.0 ± 1.42 | 100.0 ± 3.8 | 100.0 ± 1.9 | 100.0 ± 2.7 | 100.0 ± 3.5 |
| **2** | 83.7 ± 3.0 | 96.3 ± 7.2 | 104.4 ± 6.1 | 91.8 ± 3.9 | 84.6 ± 2.4 |
| **3** | 105.1 ± 7.7 | 98.0 ± 7.5 | 111.6 ± 8.6 | 98.2 ± 2.9 | 102.6 ± 4.9 |
| **4** | 91.7 ± 4.4 | 112.2 ± 6.6 | 108.9 ± 6.2 | 95.7 ± 4.3 | 99.2 ± 2.2 |
| **5** | 92.0 ± 3.5 | 106.0 ± 6.5 | 99.1 ± 4.6 | 108.8 ± 2.5 | 92.9 ± 2.3 |
| **6** | 108 ± 3.9 | 99.7 ± 4.6 | 95.6 ± 5.1 | 105.1 ± 3.5 | 93.1 ± 2.4 |
| **7** | 84.7 ± 9.9 | 101.2 ± 5.8 | 106.6 ± 7.0 | 105.6 ± 4.3 | 76.0 ± 3.6 |
| **8** | 89.9 ± 2.1 | 95.3 ± 5.9 | 100.3 ± 6.1 | 107.5 ± 3.7 | 92.9 ± 2.5 |
| **9** | 97.2 ± 2.3 | 102.4 ± 7.4 | 111.8 ± 5.2 | 97.8 ± 4.7 | 88.0 ± 3.3 |
| **10** | 101.7 ± 3.7 | 105.4 ± 5.5 | 104.2 ± 10.3 | 48.43 ± 13.7 | 47.6 ± 9.7 |

*[a]cAMP$_i$ refers to the inhibition of the forskolin-induced cAMP production.*

**Supplementary Table S2. Transduction coefficient (log(τ/KA)) values of each compound on Gα$_{i/o}$, cAMP inhibition, and β-arrestin recruitment.** Each value represents the mean of three independent experiments, which were each done in triplicate, and is expressed as the mean ± S.E.M.

| analog | cAMP$_i$[a] log(τ/K$_A$) | Gα$_{i1}$ log(τ/K$_A$) | Gα$_{oA}$ log(τ/K$_A$) | β-arrestin-1 log(τ/K$_A$) | β-arrestin-2 log(τ/K$_A$) |
|---|---|---|---|---|---|
| **1** | 8.77 ± 0.05 | 9.05 ± 0.15 | 7.98 ± 0.19 | 7.54 ± 0.05 | 7.51 ± 0.10 |
| **2** | 8.16 ± 0.10 | 8.46 ± 0.19 | 8.99 ± 0.18 | 6.93 ± 0.06 | 6.77 ± 0.13 |
| **3** | 9.45 ± 0.10 | 8.90 ± 0.17 | 9.08 ± 0.17 | 8.49 ± 0.05 | 8.02 ± 0.08 |
| **4** | 8.82 ± 0.13 | 8.60 ± 0.16 | 8.30 ± 0.17 | 6.96 ± 0.06 | 6.90 ± 0.14 |
| **5** | 9.22 ± 0.10 | 8.91 ± 0.16 | 8.89 ± 0.18 | 8.11 ± 0.04 | 7.89 ± 0.10 |
| **6** | 10.48 ± 0.08 | 8.96 ± 0.16 | 8.60 ± 0.19 | 7.83 ± 0.04 | 7.42 ± 0.12 |
| **7** | 6.68 ± 0.13 | 8.19 ± 0.18 | 8.36 ± 0.18 | 6.89 ± 0.05 | 6.43 ± 0.14 |
| **8** | 10.14 ± 0.10 | 8.43 ± 0.20 | 8.44 ± 0.18 | 8.45 ± 0.05 | 8.09 ± 0.10 |
| **9** | 8.45 ± 0.10 | 8.18 ± 0.17 | 8.03 ± 0.17 | 6.96 ± 0.06 | 6.89 ± 0.12 |
| **10** | 7.91 ± 0.09 | 8.51 ± 0.25 | 9.62 ± 0.19 | 6.30 ± 0.16 | 6.48 ± 0.25 |

[a]*cAMP$_i$ refers to the inhibition of the forskolin-induced cAMP production.*

**Supplementary Table S3. Normalized transduction coefficient ($\Delta$log($\tau$/KA)) values and adjusted *p* value for each compound on Gα$_{i/o}$, cAMP inhibition, and β-arrestin recruitment.** $\Delta$log($\tau$/KA) values are expressed as the mean ± S.E.M. Adjusted p values result from the statistical comparison of the $\Delta$log($\tau$/KA) on each pathway with a Kruskal-Wallis followed by a Dunn's post hoc test. An adjusted *p* < 0.05 was considered significant. Shaded boxes represent significantly different values compared to apelin-13.

| analog | cAMP$_i$[a] | | Gα$_{i1}$ | | Gα$_{oA}$ | | β-arrestin-1 | | β-arrestin-2 | |
|---|---|---|---|---|---|---|---|---|---|---|
| | $\Delta$log($\tau$/K$_A$) | Adjusted *p* value | $\Delta$log($\tau$/K$_A$) | Adjusted *p* value | $\Delta$log($\tau$/K$_A$) | Adjusted *p* value | $\Delta$log($\tau$/K$_A$) | Adjusted *p* value | $\Delta$log($\tau$/K$_A$) | Adjusted *p* value |
| **1** | 0.00 ± 0.07 | - | 0.00 ± 0.22 | - | 0.00 ± 0.26 | - | 0.00 ± 0.07 | - | 0.00 ± 0.14 | - |
| **2** | -0.61 ± 0.11 | 0.0002 | -0.60 ± 0.25 | 0.3146 | 1.01 ± 0.26 | 0.0349 | -0.60 ± 0.08 | 0.0001 | -0.74 ± 0.16 | 0.0072 |
| **3** | 0.68 ± 0.11 | 0.0001 | -0.16 ± 0.24 | 0.9972 | 1.09 ± 0.26 | 0.0187 | 0.95 ± 0.07 | 0.0001 | 0.51 ± 0.13 | 0.1251 |
| **4** | 0.05 ± 0.14 | 0.9994 | -0.45 ± 0.23 | 0.6345 | 0.32 ± 0.25 | 0.9410 | -0.57 ± 0.07 | 0.0001 | -0.61 ± 0.18 | 0.0408 |
| **5** | 0.45 ± 0.11 | 0.0098 | -0.14 ± 0.22 | 0.9993 | 0.91 ± 0.26 | 0.0722 | 0.57 ± 0.06 | 0.0001 | 0.38 ± 0.14 | 0.3960 |
| **6** | 1.71 ± 0.09 | 0.0001 | -0.09 ± 0.22 | 0.9996 | 0.62 ± 0.27 | 0.3877 | 0.30 ± 0.06 | 0.0521 | -0.09 ± 0.15 | 0.9994 |
| **7** | -2.09 ± 0.14 | 0.0001 | -0.86 ± 0.24 | 0.0536 | 0.38 ± 0.26 | 0.8633 | -0.64 ± 0.07 | 0.0001 | -1.08 ± 0.17 | 0.0001 |
| **8** | 1.37 ± 0.12 | 0.0001 | -0.62 ± 0.25 | 0.2810 | 0.46 ± 0.26 | 0.7133 | 0.91 ± 0.07 | 0.0001 | 0.57 ± 0.14 | 0.0653 |
| **9** | -0.32 ± 0.11 | 0.1246 | -0.87 ± 0.23 | 0.0495 | 0.04 ± 0.25 | 0.9999 | -0.58 ± 0.07 | 0.0001 | -0.62 ± 0.15 | 0.0361 |
| **10** | -0.86 ± 0.10 | 0.0001 | -0.54 ± 0.29 | 0.4302 | 1.64 ± 0.26 | 0.0001 | -1.23 ± 0.17 | 0.0001 | -1.03 ± 0.27 | 0.0001 |

[a]cAMP$_i$ refers to the inhibition of the forskolin-induced cAMP production.

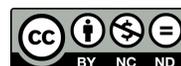